\documentstyle[12pt]{article}

\font\twlgot =eufm10 scaled \magstep1
\font\egtgot =eufm8
\font\sevgot =eufm7

\font\twlmsb =msbm10 scaled \magstep1
\font\egtmsb =msbm8
\font\sevmsb =msbm7

\newfam\gotfam
\def\pgot{\fam\gotfam\twlgot}
\textfont\gotfam\twlgot
\scriptfont\gotfam\egtgot
\scriptscriptfont\gotfam\sevgot
\def\got{\protect\pgot}

\newfam\msbfam
\textfont\msbfam\twlmsb
\scriptfont\msbfam\egtmsb
\scriptscriptfont\msbfam\sevmsb
\def\Bbb{\protect\pBbb}
\def\pBbb{\relax\ifmmode\expandafter\Bb\else\typeout{You cann't use
Bbb in text mode}\fi}
\def\Bb #1{{\fam\msbfam\relax#1}}

\def\thebibliography#1{\bigskip\section*{\large
\bf References\\}\list
  {[\arabic{enumi}]}{\settowidth\labelwidth{#1}\leftmargin\labelwidth
    \advance\leftmargin\labelsep
    \usecounter{enumi}}
    \def\newblock{\hskip .11em plus .33em minus .07em}
    \sloppy\clubpenalty4000\widowpenalty4000
    \sfcode`\.=1000\relax}

\def\op#1{\mathop{{\it\fam0} #1}\limits}

\newcommand{\beq}{\begin{equation}}
\newcommand{\eeq}{\end{equation}}
\newcommand{\ben}{\begin{eqnarray}}
\newcommand{\een}{\end{eqnarray}}
\newcommand{\be}{\begin{eqnarray*}}
\newcommand{\ee}{\end{eqnarray*}}
\newcommand{\bea}{\begin{eqalph}}
\newcommand{\eea}{\end{eqalph}}

\newcommand{\gT}{{\got T}}

\newcommand{\cL}{{\cal L}}

\newcommand{\bL}{{\bf L}}

\newcommand{\al}{\alpha}
\newcommand{\bt}{\beta}
\newcommand{\dl}{\delta}
\newcommand{\la}{\lambda}

\newcommand{\f}{\phi}

\newcommand{\x}{\xi}

\newcommand{\m}{\mu}
\newcommand{\n}{\nu}

\newcommand{\e}{\epsilon}
\newcommand{\ve}{\varepsilon}

\newcommand{\si}{\sigma}
\newcommand{\Si}{\Sigma}

\newcommand{\wt}{\widetilde}

\newcommand{\dr}{\partial}

\newcommand{\ot}{\otimes}
\newcommand{\ap}{\approx}

\newenvironment{eqalph}{\stepcounter{equation}
\setcounter{equationa}{\value{equation}}
\setcounter{equation}{0}

\begin{eqnarray}}{\end{eqnarray}\setcounter{equation}{\value{equationa}}}

\newcommand{\mar}[1]{}

\begin{document}
\hbox{}

{\parindent=0pt

{\large \bf The gauge condition in gravitation theory
with a background metric}
\bigskip 

{\bf G. A. Sardanashvily}

\medskip

\begin{small}

Department of Theoretical Physics, Moscow State University, 117234
Moscow, Russia

E-mail: sard@grav.phys.msu.su

URL: http://webcenter.ru/$\sim$sardan/
\bigskip

{\bf Abstract.}
In gravitation theory with a background metric, a gravitational field
is described by a two-tensor field. The energy-momentum
conservation law imposes a gauge condition on this field.
\end{small}
}

\bigskip
\bigskip

Gravitation theory in the presence of a background metric remains under
consideration. In particular, there are two variants of gauge
gravitation theory \cite{tmf,sard98a,tmf2}. The first of them leads to 
the metric-affine gravitation theory, while 
the second one (henceforth BMT) deals with a
background pseudo-Riemannian metric $g^{\m\n}$ and a non-degenerate 
(1,1)-tensor field $q^\m{}_\nu$,
regarded as a gravitational field. A Lagrangian of BMT is of the form
\mar{b2}\beq
\cL_{\rm BMT}=  \e \cL_q+ \cL_{\rm AM}+\cL_{\rm
m},  \label{b2}
\eeq 
where $\cL_q$ is a Lagrangian of a tensor gravitational field
$q^\m{}_\n$ in the presence of a background metric $g^{\m\n}$,
$\cL_{\rm AM}$ is a Lagrangian of the metric-affine theory where
a metric $g^{\m\n}$ is replaced with an effective metric
\mar{b3}\beq
\wt g^{\m\n}=q^\m{}_\al q^\n{}_\bt g^{\al\bt}, \label{b3}
\eeq
and $\cL_{\rm m}$ is a matter field Lagrangian depending on an
effective metric $\wt g$ and a general linear connection $K$ 
(see, e.g., \cite{log}).
Note that, strictly speaking, $\wt g$ (\ref{b3}) is not a metric, but
there exists a metric whose coefficients equal $\wt g^{\m\n}$ (\ref{b3}).
Therefore, one usually assumes that $\cL_q$ depends on $q$ only via
an effective metric $\wt g$.

A glance at the expression (\ref{b2}) shows that the
matter field equation in BMT is that of affine-metric theory where a
metric $g$ is replaced with an effective metric $\wt g$. However,
gravitational field equations are different because of the term $\e\cL_q$. 
The question is whether solutions of BMT come to solutions of the
metric-affine theory if the constant 
$\e$ tends to zero. The answer to this question follows from the
energy-momentum conservation law in BMT. We obtain the weak equality
\mar{b4}\beq
\nabla_\la t^\la_\al\ap 0, \label{b4}
\eeq
where $\nabla_\la$ is the covariant derivative with respect to the
Levi-Civita connection of an effective metric $\wt g$ and $t^\la_\al$ is the
metric energy-momentum tensor of the Lagrangian $\cL_q$ with respect to
$\wt g$. The equality
(\ref{b4}) holds for any solution $q$ of field equations of 
BMT. This equality is defined only by the Lagrangian $\cL_q$, and 
is independent of other fields and the constant $\e$. 
Therefore, it can be regarded as a gauge condition on solutions of BMT.

Recall that, in gauge theory on a
fibre bundle $Y\to X$ coordinated by $(x^\la,y^i)$,
gauge
transformations are defined as bundle automorphisms of $Y\to X$
(see \cite{book,book00,epr02} for a survey).
Their infinitesimal
generators are projectable vector fields 
\mar{e1}\beq
u=u^\la(x^\m)\dr_\la +u^i(x^\m,y^j)\dr_i \label{e1}
\eeq
on a fibre bundle $Y\to X$. 
We are concerned with a first order Lagrangian field theory on $Y$. 
Its configuration space is the
first order jet manifold $J^1Y$ of $Y\to X$ coordinated by
$(x^\la,y^i,y^i_\la)$. A first order Lagrangian
is defined as a density
\mar{cc201}\beq
L=\cL(x^\la,y^i,y^i_\la)dx^n, \qquad
n=\dim X, \label{cc201}
\eeq
on $J^1Y$. A Lagrangian $L$ is invariant under
a one-parameter group of gauge transformations generated by a vector
field $u$ (\ref{e1}) iff its Lie derivative 
\mar{e100}\beq
\bL_{J^1u}L=J^1u\rfloor dL +d(J^1u\rfloor L) \label{e100}
\eeq
along the jet
prolongation $J^1u$ of $u$ onto $J^1Y$
vanishes. By virtue of the well-known first variation
formula, the Lie derivative (\ref{e100}) admits the canonical decomposition
\mar{b20}\beq
\bL_{J^1u}L=(u^i-y^i_\m u^\m)\dl_i\cL d^nx-d_\la\gT^\la d^nx, \label{b20}
\eeq
where $d_\la=\dr_\la +y^i_\la\dr_i +y_{\la\m}\dr^\m_i$ denotes the
total derivative, 
\mar{b21}\beq
\dl_i\cL=(\dr_i -d_\al\dr^\la_i)\cL \label{b21}
\eeq
are variational derivatives, and 
\mar{Q30}\beq
\gT^\la =(u^\m y^i_\m-u^i)\dr^\la_i\cL -u^\la\cL
\label{Q30}
\eeq
is a current along the vector field $u$. If the Lie derivative
$\bL_{J^1u}L$ vanishes, the first variation formula 
on the shell $\dl_i\cL=0$ leads to the weak conservation law 
$d_\la\gT^\la_u\ap 0$.
In particular, if $u=\tau^\la\dr_\la + u^i\dr_i$ is a 
lift onto $Y$ of a vector field
$\tau=\tau^\la\dr_\la$ on $X$ (i.e., $u^i$ is linear in $\tau^\la$ and
their derivatives), then $\gT^\la$ is an energy-momentum current
\cite{fer,book,got,sard97}. Note that  
different lifts onto $Y$ of a vector field
$\tau$ on $X$ lead to distinct energy-momentum currents 
whose difference is a Noether current.  Gravitation theory 
deals with fibre
bundles over $X$ which admit the canonical lift of any vector field
on $X$. This lift is a generator of general covariant transformations 
\cite{book,sard97b}.

Let $X$ be a 4-dimensional oriented smooth manifold satisfying the well-known
topological conditions of the existence of a pseudo-Riemannian metric. 
Let $LX$ denote the fiber bundle of
oriented frames in $TX$. It is a principal bundle with the structure
group $GL_4=GL^+(4,\Bbb R)$.
A pseudo-Riemannian metric $g$ on $X$ is defined as a global
section of the quotient  bundle
\mar{b10}\beq
\Si=LX/SO(1,3)\to X. \label{b10}
\eeq
This bundle is identified with an open subbundle of the tensor bundle
$\op\vee^2 TX$. Therefore, $\Si$ can be equipped with
coordinates $(\x^\la,\si^{\m\n})$, and $g$ is represented by tensor
fields $g^{\m\n}$ or $g_{\m\n}$. 

Principal connections $K$ on the frame bundle $LX$ are linear connections
\mar{B}\beq
K= dx^\la\otimes (\dr_\la +K_\la{}^\m{}_\n \dot x^\n
\dot\dr_\m) \label{B}
\eeq
on the tangent bundle $TX$ and other tensor bundles over $X$. They 
are represented by sections of the quotient bundle 
\mar{015}\beq
C=J^1LX/GL_4\to X, \label{015}
\eeq
where $J^1LX$ is the first order jet manifold of the frame bundle
$LX\to X$ \cite{book,book00,epr02}.
The bundle of connections $C$ is equipped with the coordinates
$(x^\la, k_\la{}^\nu{}_\al)$ such that the coordinates
$k_\la{}^\nu{}_\al\circ K=K_\la{}^\nu{}_\al$ of any section $K$
are the coefficients of corresponding linear connection (\ref{B}). 

A tensor gravitational field $q$ is defined as a section of the
group bundle $Q\to X$ associated with $LX$. Its typical fiber is the group
$GL_4$ which acts on itself by the adjoint representation.
The group bundle $Q$ as a subbundle of the tensor bundle $TX\ot
T^*X$ is equipped with the coordinates
$(x^\la, q^\la{}_\m)$.
The canonical left action $Q$ on any bundle associated with $LX$ is
given. In particular, its action on the quotient bundle $\Si$
(\ref{b10}) takes the form 
\be
\rho: Q\times \Si\to \Si, \qquad 
\rho:(q^\la{}_\m, \si^{\m\n})\mapsto 
\wt \si^{\m\n}=q^\m{}_\al q^\n{}_\bt \si^{\al\bt}.
\ee

Since the Lagrangian (\ref{b2}) of BMT depends on $q$
only via an effective metric, let us further replace variables $q$ with
the variables $\wt \si$. 
Then, the configuration space of BMT is the jet manifold $J^1Y$ of the product
$Y=\Si\times C\times Z$,
where $Z\to X$ is a fibre bundle of
matter fields. Relative to coordinates $(\wt \si^{\m\n},
k_\la{}^\al{}_\bt, \f)$ on $Y$,  the Lagrangian (\ref{b2}) reads
\mar{b13}\beq
\cL_{\rm BMT}= \e\cL_q(\si,\wt\si) +\cL_{\rm AM}(\wt\si,R)
+\cL_m(\f,\wt\si,k),  \label{b13}
\eeq
where the metric-affine Lagrangian $\cL_{AM}$ is expressed into
components of the curvature tensor
\be
R_{\la\m}{}^\al{}_\bt=k_{\la\m}{}^\al{}_\bt -k_{\m\la}{}^\al{}_\bt
+k_\la{}^\ve{}_\bt k_\m{}^\al{}_\ve - k_\m{}^\ve{}_\bt k_\la{}^\al{}_\ve
\ee
contracted by means of the effective
metric $\wt\si$. The corresponding field equations read
\mar{b31-3}\ben
&& \dl_{\m\n}(\e\cL_q + \cL_{\rm AM}+ \cL_m)=0, \label{b31}\\
&& \dl^\la{}_\al{}^\bt (\cL_{\rm AM}+ \cL_m)=0, \label{b32}\\
&& \dl_\f\cL_m=0. \nonumber
\een
where $\dl_{\m\n}$, $\dl^\la{}_\al{}^\bt$ and $\dl_\f$ are variational
derivatives (\ref{b21}) wit respect to $\wt \si^{\m\n}$,
$k_\la{}^\al{}_\bt$ and $\f$. 

Let $\tau=\tau^\la\dr_\la$ be a vector field on $X$. Its canonical lift
onto the fibre bundle $\Si\times C$ reads
\mar{051}\ben
\wt\tau = \tau^\la\dr_\la +
&& (\dr_\nu\tau^\al k_\m{}^\nu{}_\bt - \dr_\bt\tau^\nu
k_\m{}^\al{}_\nu - \dr_\m\tau^\nu
k_\nu{}^\al{}_\bt +\dr_{\m\bt}\tau^\al)\frac{\dr}{\dr k_\m{}^\al{}_\bt}
\label{051}\\
&& \qquad + (\dr_\ve\tau^\m \wt \si^{\ve\nu} + 
\dr_\ve\tau^\nu \wt \si^{\m\ve}) \frac{\dr}{\dr \wt\si^{\m\nu}} =
\tau^\la\dr_\la + u_\m{}^\al{}_\bt\dr^\m{}_\al{}^\bt + u^{\m\nu}\dr_{\m\nu}.
\nonumber
\een
It is a generator of general covariant transformations of the fiber
bundle $\Si\times C$.
Let us apply the first variation formula (\ref{b20}) to the Lie
derivative $\bL_{J^1\wt\tau}L_{\rm MA}$. Since the Lagrangian 
$L_{\rm MA}$ is invariant under general covariant transformations, we obtain
the equality
\mar{b40}\beq
0=(u^{\m\n}-\si_\ve^{\m\n}\tau^\ve)\dl_{\m\n}\cL_{\rm MA} +
(u_\la{}^\al{}_\bt - k_{\ve\la}{}^\al{}_\bt\tau^\ve)\dl^\la{}_\al{}^\bt
\cL_{\rm MA} - d_\la\gT^\la_{\rm MA}, \label{b40}
\eeq
where $\gT_{\rm MA}$ is the energy-momentum current of the metric-affine 
gravitation theory. It reads
\mar{b41}\beq
\gT^\la_{\rm MA}=2\wt\si^{\la\m}\tau^\al\dl_{\al\m}\cL_{\rm MA} + 
T(\dl^\la{}_\al{}^\bt\cL_{\rm MA}) + d_\m U_{\rm MA}^{\m\la}, \label{b41}
\eeq
where $T(.)$ are terms linear in the variational derivatives
$\dl^\la{}_\al{}^\bt\cL_{\rm MA}$ and $U_{\rm MA}^{\m\la}$ is the
generalized Komar superpotential \cite{giach,book}.

For the sake of simplicity, let us assume that there exists a domain
$N\subset X$ where $L_m=0$, and let us consider the equality (\ref{b40})
on $N$. The field equations (\ref{b31}) -- (\ref{b32}) on $N$ take the form
\mar{b42,3}\ben
&& \dl_{\m\n}(\e\cL_q + \cL_{\rm AM})=0, \label{b42}\\
&& \dl^\la{}_\al{}^\bt \cL_{\rm AM}=0, \label{b43}
\een
and the energy current (\ref{b41}) reads
\mar{b45}\beq
\gT^\la_{\rm MA}=2\wt\si^{\la\m}\tau^\al\dl_{\al\m}\cL_{\rm MA} + 
d_\m U_{\rm MA}^{\m\la}. \label{b45}
\eeq
Substituting (\ref{b42}) and (\ref{b45}) into (\ref{b40}), we obtain
the weak equality
\mar{b46}\beq
0\ap
-(2\dr_\la\tau^\al\si^{\la\m}-\si_\la^{\al\m}\tau^\la)\dl_{\al\m}\cL_q
+ d_\la 
(2\wt\si^{\la\m}\tau^\al\dl_{\al\m}\cL_q). \label{b46}
\eeq
A simple computation brings this equality into the desired form
(\ref{b4}) where
\be
t^\la_\al=2\wt g^{\la\n}\sqrt{-\wt g}\dl_{\n\al}\cL_q.
\ee

\end{document}